\documentclass[10pt,twocolumn,letterpaper]{article}

\usepackage{cvpr}
\usepackage{times}
\usepackage{epsfig}
\usepackage{graphicx}
\usepackage{amsmath}
\usepackage{amssymb}
\usepackage{subfigure}
\usepackage{color}
\usepackage{setspace}
\usepackage{algorithm}
\usepackage{algorithmic}
\usepackage{multirow}
\usepackage[L7x,T1]{fontenc}
\usepackage[utf8]{inputenc}
\usepackage[lithuanian,english]{babel}

\usepackage[pagebackref=true,breaklinks=true,letterpaper=true,colorlinks,bookmarks=false]{hyperref}

\cvprfinalcopy 

\def\cvprPaperID{1736} 

\ifcvprfinal\fi
\begin{document}

\title{Cross-lingual Information Retrieval with BERT}

\author{Zhuolin Jiang$^{\dag}$, Amro El-Jaroudi$^{\dag}$$^{\ddag}$, William Hartmann$^{\dag}$, Damianos Karakos$^{\dag}$, Lingjun Zhao$^{\dag}$\\
$^\dag$Raytheon BBN Technologies, Cambridge, MA, 02138\\
$^\ddag$University of Pittsburgh, Pittsburgh, PA, 15261\\
{\tt\small \{zhuolin.jiang, amro.a.eljaroudi-nr, william.hartmann, damianos.karakos, lingjun.zhao\}@raytheon.com}
}

\maketitle

\begin{abstract}
Multiple neural language models have been developed recently, \textit{e.g.}, BERT and XLNet, and achieved impressive results in various NLP tasks including sentence classification, question answering and document ranking. In this paper, we explore the use of the popular bidirectional language model, BERT, to model and learn the relevance between English queries and foreign-language documents in the task of cross-lingual information retrieval. A deep relevance matching model based on BERT is introduced and trained by finetuning a pretrained multilingual BERT model with weak supervision, using home-made CLIR training data derived from parallel corpora. Experimental results of the retrieval of Lithuanian documents against short English queries show that our model is effective and outperforms the competitive baseline approaches. 
\end{abstract}

\section{Introduction}

A traditional cross-lingual information retrieval (CLIR) system consists of two components: machine translation and monolingual information retrieval~\cite{nie2010}. The idea is to solve the translation problem first, then the cross-lingual IR problem become monolingual IR. However, the performance of translation-based approaches is limited by the quality of the machine translation and it needs to handle to translation ambiguity~\cite{Zhou2012}. One possible solution is to consider the translation alternatives of individual words of queries or documents as in~\cite{zbib19,XuEMNLP00}, which provides more possibilities for matching query words in relevant documents compared to using single translations. But the alignment information is necessarily required in the training stage of  the CLIR system to extract target-source word pairs from parallel data and this is not a trivial task.

To achieve good performance in IR, deep neural networks have been widely used in this task. These approaches can be roughly divided into two categories. The first class of approaches uses pretrained word representations or embeddings, such as word2vec~\cite{tomas2013} and GloVe~\cite{pennington2014glove}, directly to improve IR models. Usually these word embeddings are pretrained on large scale text corpora using co-occurrence statistics, so they have modeled the underlying data distribution implicitly and should be helpful for building discriminative models.~\cite{VulicSIGIR15} and ~\cite{LitschkoSIGIR18} used pretrained bilingual embeddings to represent queries and foreign documents, and then ranked documents by cosine similarity. ~\cite{zheng2015} used word2vec embeddings to learn query term weights. However, their training objectives of trained neural embeddings are different from the objective of IR.

The second set of approaches design and train deep neural networks based on IR objectives. These methods have shown impressive results on monolingual IR datasets~\cite{XiongSIGIR17,guocikm16,mostafaSIGIR17}.  They usually rely on  large amounts of query-document relevance annotated data that are expensive to obtain, especially for low-resource language pairs in cross-lingual IR tasks. Moreover, it is not clear whether they generalize well when documents and queries are in different languages.

Recently multiple pretrained language models have been developed such as BERT~\cite{devlin-etal-2019-bert} and XLNet~\cite{xlnet2019}, that model the underlying data distribution and learn the linguistic patterns or features in language. These models have outperformed traditional word embeddings on various NLP tasks~\cite{xlnet2019,devlin-etal-2019-bert,peters-etal-2018-deep,Lan2019ALBERTAL}. These pretrained models also provided new opportunities for IR. Therefore, several recent works have successfully applied BERT pretrained models for monolingual IR~\cite{daiSIGIR19,akkalyoncu-yilmaz-etal-2019-applying} and passage re-ranking~\cite{Nogueira19}.

In this paper, we extend and apply BERT as a ranker for CLIR. We introduce a cross-lingual deep relevance matching model for CLIR based on BERT. We finetune a pretrained multilingual model with home-made CLIR data and obtain very promising results. In order to finetune the model, we construct a large amount of training data from parallel data, which is mainly used for machine translation and is much easier to obtain compared to the relevance labels of query-document pairs. In addition, we don't require the source-target alignment information to construct training samples and avoid the quality issues of machine translation in traditional CLIR. The entire  model is specifically optimized using a CLIR objective. Our main contributions are:

\begin{itemize}
\item We introduce a cross-lingual deep relevance architecture with BERT, where a pretrained multilingual BERT model is adapted for cross-lingual IR.
\item We define a proxy CLIR task which can be used to easily construct CLIR training data from bitext data, without requiring any amount of relevance labels of query-document pairs in different languages.
\end{itemize}

\section{Our approach}
\subsection{Motivation}
\begin{figure}
  \begin{center}
    \includegraphics[width=1.0\linewidth]{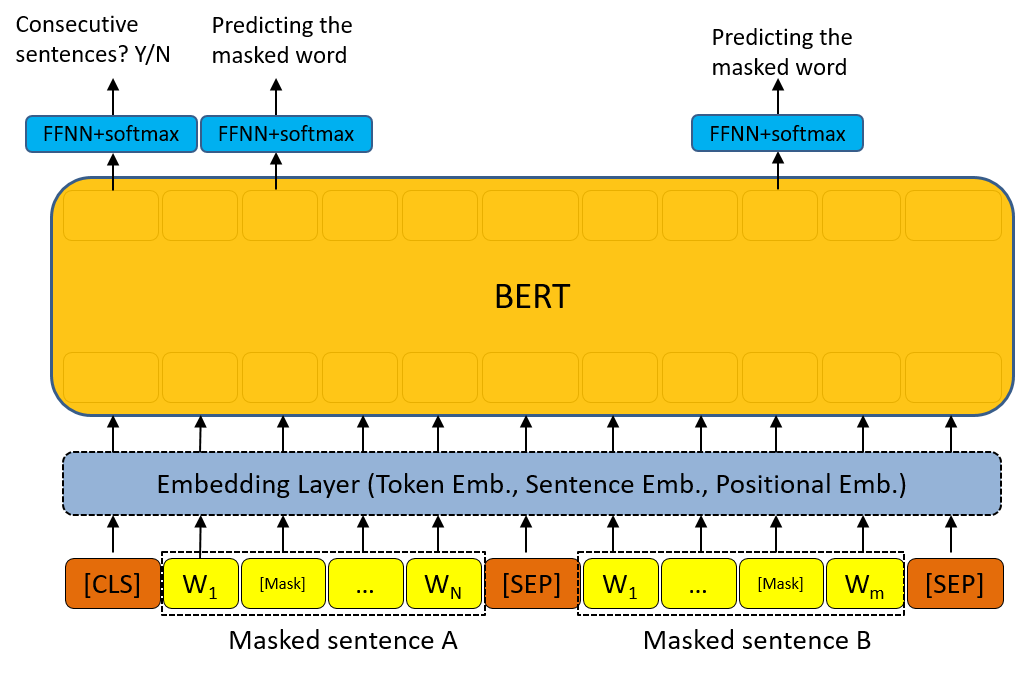}
  \end{center}
  \caption{BERT pretraining architecture \protect\cite{devlin-etal-2019-bert}. FFNN denotes feed-forward neural network.}
  \label{pretraining_architecture}
\end{figure}

BERT~\cite{devlin-etal-2019-bert}  is the first bidirectional language model, which makes use of left and right word contexts simultaneously to predict word tokens. It is trained by optimizing two objectives: masked word prediction and next sentence prediction. As shown in Figure~\ref{pretraining_architecture}, the inputs are a pair of masked sentences in the same language, where some tokens  in the both sentences are replaced by symbol `[Mask]'. The BERT model is trained to predict these masked tokens, by capturing within or across sentence meaning (or context), which is important for IR. The second objective aims to judge whether the sentences are consecutive or not. It encourages the BERT model to model the relationship between two sentences.  The self-attention mechanism in BERT models the local interactions of words in sentence A with words in sentence B, so it can learn pairwise sentence or word-token relevance patterns.  The entire BERT model is pretrained on large scale text corpora and learns linguistic patterns in language. So search tasks with little training data can still benefit from the pretrained model.

Finetuning BERT on search task makes it learn IR specific features. It can capture query-document exact term matching, bi-gram features for monolingual IR as introduced in~\cite{daiSIGIR19}. Local matchings of words and n-grams have proven to be strong neural IR features. Bigram modeling is important, because it can learn the meaning of word compounds (bi-grams) from the meanings of individual words. Motivated by this work, we aim to finetune the pretrained BERT model for cross-lingual IR.

\subsection{Finetuning BERT for CLIR}
\begin{figure}
  \begin{center}
    \includegraphics[width=0.92\linewidth]{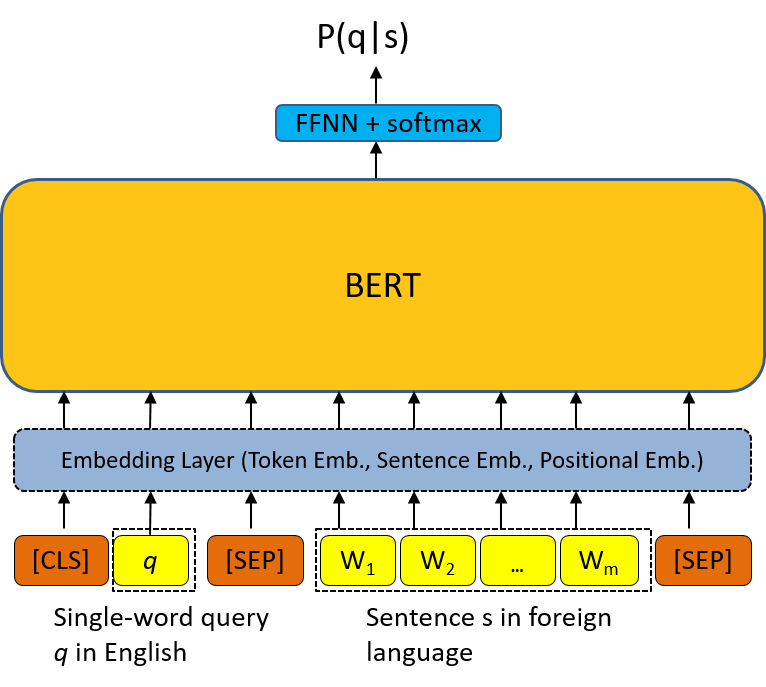}
  \end{center}
  \caption{Fine-tuned CLIR BERT model architecture. }
  \label{finetuned_architecture}
  \vspace{-0.5cm}
\end{figure}

Figure~\ref{finetuned_architecture} shows the proposed CLIR model architecture with BERT. The inputs are pairs of single-word queries $q$ in English and foreign-language sentences $s$. This is different from the pretraining model in Figure~\ref{pretraining_architecture}, where the model is fed with pairs of sentences in the same language. We concatenate the query $q$ and the foreign-language sentence $s$ into a text sequence `[[CLS], $q$, [SEP], $s$, [SEP]]'. The output embedding of the first token `[CLS]` is used as a representation of the entire query-sentence pair. Then it is fed into a single layer feed-forward neural network to predict the relevance score, which is the probability, $p(q|s)$,  of query $q$ occurring in sentence $s$.

There are three types of parameterized layers in this model: (1) an embedding layer including token embedding, sentence embedding and positional embedding~\cite{devlin-etal-2019-bert}; (2) BERT layers which are 12 layers of transformer blocks; (3) a feed-forward neural network (FFNN) which is a single layer neural network  in our implementation. The embedding layer and BERT layer are initialized with the pretrained BERT model~\footnote{We used the pretrained multi-lingual BERT model, which is trained on the concatenation of monolingual Wikipedia corpora from 104 languages. It has $12$ layers, $768$ hidden dimensions, $12$ self-attention heads and $110$ million parameters.}, while the FFNN is learned from scratch. During finetuning, the entire model is tuned to learn more CLIR-specific features. We only train the model  using single-word queries since the queries in MATERIAL dataset are typically short and keyword based,  but our approach can be easily extended to be multi-word queries or query phrases.  After finetuning, this model produces a sentence-level relevance score for a pair of input query and foreign language sentence.

For the CLIR task, given a user-issued query $Q$, the foreign-language document $Doc$ is ranked by its relevance score with respect to $Q$. The document-level relevance score $P(Doc \text{ is } R|Q)$ is calculated by aggregating the sentence-level scores with a Noisy-OR model:

\begin{align}
P(Doc \text{ is } R|Q) &= P(Q \text{ occurs at least in one sentence }  \text{ in } Doc) \nonumber \\
                        &= 1-\prod_{s \in Doc}(1-P(Q|s))  \\
                        &= 1-\prod_{s \in Doc}(1-\prod_{q \in Q}p(q|s)) \nonumber
\end{align}

Note that a multi-word query will be split into multiple single-word queries when computing document-level relevance scores. The individual query terms $q\in Q$ are modeled independently.

\begin{table*}
\begin{center}
\scalebox{0.9}{
\begin{tabular}{|p{3cm}|p{8cm}|p{2cm}|}
        \hline
    Query in English & Foreign-language sentence & Relevant \\
        \hline
doctors & medikų teigimu dabar veikianti sistema efektyvi & Yes \\
allege &  medikų teigimu dabar veikianti sistema efektyvi & Yes \\
controller & medikų teigimu dabar veikianti sistema efektyvi & No \\
leisure & medikų teigimu dabar veikianti sistema efektyvi  & No \\
 \hline
\end{tabular}}
\end{center}
\caption{Four training examples derived from a bitext: \textit{Source-Lithuanian}: medikų teigimu dabar veikianti sistema efektyvi; \textit{Target-English}:~\underline{doctors} \underline{allege} that the system currently in operation is effective. }
\label{training-example}
\end{table*}

\subsection{Finetuning using Weak Supervision}
To finetune the BERT CLIR model, we start with bitext data in English and the desired foreign-language. We then define a proxy CLIR task to construct training samples: Given a foreign-language sentence $s$ and an English query term $q$,  sentence $s$ is relevant to $q$ if $q$ occurs in one plausible translation of $s$.  Any non-stop English word in the bitext can serve as a single-word query. The English word and its the corresponding foreign-language sentence constitute a positive example. Similarly, we randomly select other words from the English vocabulary, which are not in the English sentence, to be query words to construct negative examples. Table~\ref{training-example} shows an illustration of constructing four training examples from a bitext in Lithuanian and English.  We select `doctors' and `allege' in the English sentence as two single-word queries and use the Lithuanian sentence to construct two positive examples, and pick another two words “controller” and “leisure” in the English vocabulary, which are not in the English sentence, to construct negative examples. In this way, we can construct a large-scale training corpus for CLIR using parallel data only, which are much easier to obtain compared to query-document relevance annotated data.

\section{Experiments}
We report experimental results on the retrieval of Lithuanian text and speech documents against short English queries. We use queries and retrieval corpora provided by the IARPA MATERIAL program. The retrieval corpora have two datasets: an analysis set (about 800 documents) and a development set (about 400 documents). The query set $Q1$ contains $300$ queries.

To construct the training set, we use parallel sentences released under the MATERIAL~\cite{MATERIAL} and the LORILEI~\cite{LORELEI} programs. We also include a parallel lexicon downloaded from Panlex~\cite{kamholz-etal-2014-panlex}. These parallel data contain about $2.6$ million pairs of bitexts. We extract about $54$ million training samples from these parallel data to finetune BERT. The positive-negative ratio of  CLIR training data is $1:2$.  To finetune BERT, we use the ADAM optimizer with an initial learning rate set to $1\times10^{-5}$, batch size of $32$ and max sequence length of $128$. We report the results from the model trained for one epoch.The training took one week using a Telsa V100 GPU.

We also extract $877$K testing samples from the bitexts in MATERIAL Lithuanian analysis set to test the classification accuracy of different neural CLIR models. The positive-negative ratio of this test set is $1:1$.  In addition, we evaluate our model on the MATERIAL Lithuanian analysis set and development set in terms of Mean Average Precision ( MAP) and Maximum Query Weighted Value (MQWV) scores. MQWV is used in the MATERIAL program and denotes the maximum of the metric Average Query Weighted Value (AQWV): $AQWV=1-P_{Miss} - \beta P_{FA}$, where $P_{Miss}$ is the average per-query miss rate, $P_{FA}$ is the average per-query false alarm rate and $\beta$ is a constant that changes the relative importance of the two types of error. We use $\beta = 40$. AQWV is the score using a single selected detection threshold. MQWV is the score that could be obtained with the optimal detection threshold.

To verify the effectiveness of our BERT CLIR model, we compare against four baselines:

\textbf{Probabilistic CLIR Model}~\cite{XuEMNLP00} is a generative probabilistic model which requires a probabilistic translation dictionary. The translation dictionary is generated from the word alignments of the parallel data. We used the GIZA++~\cite{och-ney-2003-systematic} and the Berkeley aligner~\cite{Haghighi09} to estimate lexical translation probabilities.

\textbf{Probabilistic Occurrence Model}~\cite{zbib19} computes the document relevance score as the probability that each query term $q$ occurs at least once in the document. $P(Doc \text{ is } R|Q) = \prod_{q\in Q}\left[1-\prod_{f \in Doc}(1-p(q|f))\right]$, where $f$ is a foreign term in the document.

\textbf{Query Relevance Attentional Neural Network Model} (QRANN)~\cite{zhao-etal-2019-weakly} uses an attention mechanism to compute a context vector derived from word embeddings in the foreign sentences, followed by a feed-forward layer to capture the relationship between query words. The idea is similar to a single transformer layer. The QRANN models are trained on multi-word queries, which are noun phrases in the English sentences of bitexts, and single-word queries.

\textbf{Dot-product Model} is a simplified version of QRANN, that computes a context vector from the word embeddings of foreign sentence using multiplicative attention, followed by the dot product of between the query embeddings and the context vector. The dot-product model is trained using single-word queries only.

\subsection{Classification Accuracy of different neural CLIR models}
The QRANN and Dot-product models are trained using the same CLIR training data used to train BERT model described earlier. The classification results of different neural CLIR approaches are shown in Table~\ref{classification_accuracy}. The CLIR BERT model achieves the best result compared to  other two neural models. From the confusion matrix in the table, BERT significantly improves the performance of classifying relevant query-sentence pairs (\textit{i.e.}, true positives), while matching the performance of classifying irrelevant query-sentence pairs (\textit{i.e.}, true negatives).

\begin{table}[t]
\centering
\begin{tabular}{|c|c|c|c|}
\hline
Approach & Accuracy &  \multicolumn{2}{c|}{Confusion Matrix}  \\
\hline
 \multirow{2}{*}{BERT}&   \multirow{2}{*}{\textbf{95.3\%}}  & 0.93 & 0.07  \\
 \cline{3-4}
 & & 0.02 & 0.98 \\
\hline
 \multirow{2}{*}{Dot-Product}&   \multirow{2}{*}{84.2\%}  & 0.74 & 0.26  \\
 \cline{3-4}
 & & 0.07 & 0.93 \\
\hline
 \multirow{2}{*}{QRANN}&   \multirow{2}{*}{87.3\%}  & 0.73 & 0.27  \\
 \cline{3-4}
 & & 0.003 & 0.997 \\
\hline
\end{tabular}
\vspace{0.2cm}
\caption{Performance of classification accuracy on the generated query-sentence pairs from the bitexts of the  MATERIAL analysis set. The first column in the confusion matrix corresponds to the positive class (\textit{i.e.}, relevant query-sentence pair) while the second the column is the negative class.}
\label{classification_accuracy}
\end{table}

\begin{table}
\centering
\begin{tabular}{|c|c|c|c|}
\hline
Approach & phrase query subset  &  entire query set  \\
\hline
Prob. CLIR & 57.4 & \textbf{61.2} \\
Prob. Occurrence & 51.4 & 56.9 \\
BERT & \textbf{61.3} & 56.8 \\
Dot-Product & 50.8 & 39.2 \\
QRANN & 55.8 & 45.5 \\
\hline
\end{tabular}
\vspace{0.2cm}
\caption{Performance of MAP scores on the MATERIAL analysis set and Q1 queries.}
\label{map_scores}
\end{table}

\begin{figure*}
  \begin{center}
    \includegraphics[width=1.0\linewidth]{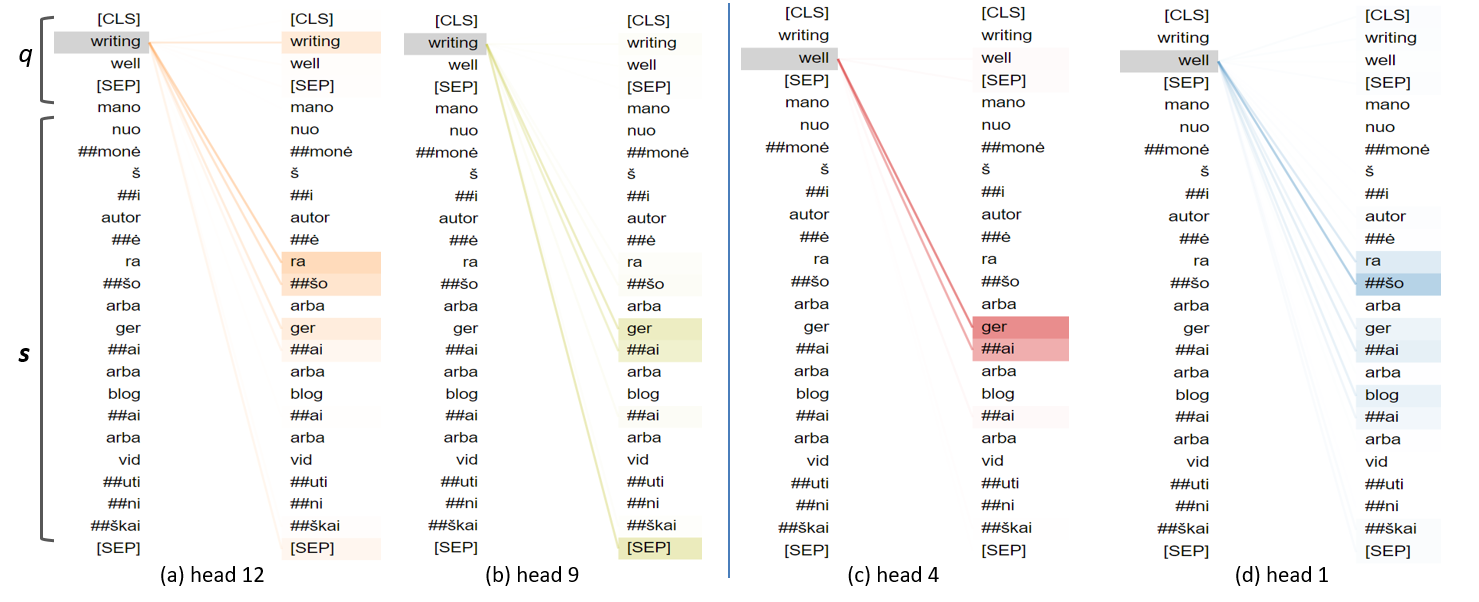}
  \end{center}
  \caption{Visualization of CLIR BERT model. Colors identify the corresponding attention heads, while the line weight reflects the attention score. Different heads from layer $12$  can capture different matching features. Word pieces` ra' , `\#\#šo' in Lithuanian correspond to `'write' in English while `ger', `\#\#ai' are for `well' in English. Head 12 and head 4 in (a)(c) can capture source-target word matching, head9 and head1 in (b)(d) could attend to its previous or next words (bigram modeling).  }
  \label{neural_head}
\end{figure*}

\subsection{MAP scores of different CLIR models}
We compare the MAP score of the BERT model with those of other CLIR models in Table~\ref{map_scores}. In the table, we report MAP scores on the phrase query subset and the entire query set separately, to see how our model trained with single-word queries performs on query phrases. In the model training stage, QRANN model is the only model that is trained with the query phrases directly, all other models (including BERT) in this experiment will split a multi-word query or query phrase into multiple single-word queries. Surprisingly, the BERT MAP scores for the phrase query subset is the best compared with the performances of other approaches. It shows that BERT model can produce  better relevance model for single-word queries and foreign-language sentence.The table also shows that BERT outperforms the other neural approaches over the entire query set.

\subsection{MQWV scores of different CLIR models}
We compare BERT models with other CLIR models in terms of MQWV scores. The results are summarized in Table~\ref{mqwv_scores}. The first row in the table shows the best results of non-neural CLIR models, which are probabilistic CLIR model and probabilistic occurrence model. In this table, we separate the results based on the type of source documents: text or speech. Speech documents are converted into text documents via automatic speech recognition~\cite{Povey_ASRU2011_2011}. The results of the BERT model on the speech sets are the best, compared with the non-neural CLIR systems, QRANN and Dot-product models,  while the results on the text sets are comparable to those from the non-neural systems, and better than the other neural systems.

\begin{table}
\centering
\begin{tabular}{|c|c|c|c|c|}
\hline
\multirow{2}{*}{Approach} &  \multicolumn{2}{c|}{Analysis Set} &  \multicolumn{2}{c|}{Development Set}  \\
\cline{2-5}
& Text& Speech & Text & Speech \\
\hline
Best non-neural system &\textbf{ 66.3} & 63.3 & \textbf{68.8} & 64.0 \\
\hline
BERT & 65.7 & \textbf{65.4} & 61.8 & \textbf{65.1} \\
\hline
Dot-Product & 61.0 & 60.4 & 56.1 & 63.7 \\
\hline
QRANN & 62.3 & 58.4 & 57.2 & 65.0 \\
\hline
\end{tabular}
\vspace{0.2cm}
\caption{MQWV scores  on the Lithuanian analysis and development sets and Q1 queries.}
\label{mqwv_scores}
\end{table}

\subsection{Analysis on attention patterns from BERT}
In Figure~\ref{neural_head}, we visualize the attention patterns produced by the attention heads from a transformer layer for the input English query `writing well' and the foreign-language sentence `mano nuomone ši autore rašo arba gerai arba blogai arba vidutiniškai'. The query term `writing' attends to the foreign word `rašo' (source-target word matching), while also attends to the foreign word `gerai' , which correspond to the next English word `well' in the query (bigram modeling). BERT CLIR model is able to capture these local matching features, which have been proven to be strong neural IR features.

\section{Conclusions}
We introduce a deep relevance matching model based on  BERT language modeling architecture for cross-lingual document retrieval. The self-attention based architecture models the interactions of query words with words in the foreign-language sentence.  The relevance model is initialized by the pretrained multi-lingual BERT model, and then finetuned with home-made CLIR training data that are derived from parallel data. The results of the CLIR BERT model on the data released by the MATERIAL program are better than two other competitive neural baselines, and comparable to the results of the probabilistic CLIR model. Our future work will use public IR datasets in English to learn IR features with BERT and transfer them to cross-lingual IR.

\section*{Acknowledgement}
This work was supported by the Intelligence Advanced Research Projects Activity (IARPA) via Department of Defense US Air Force Research Laboratory contract number FA8650-17-C-9118.

\section{Bibliographical References}
\label{reference}

{\small
\bibliographystyle{ieee}
\bibliography{egbib}
}

\end{document}